# High-Rate Full-Diversity Space-Time Block Codes for Three and Four Transmit Antennas[*]

Ertuğrul Başar and Ümit Aygölü

Istanbul Technical University, Faculty of Electrical & Electronics Engineering,
34469, Maslak, Istanbul, Turkey

{basarer,aygolu}@itu.edu.tr

**Abstract**

In this paper, we deal with the design of high-rate, full-diversity, low maximum likelihood (ML) decoding complexity space-time block codes (STBCs) with code rates of 2 and 1.5 complex symbols per channel use for multiple-input multiple output (MIMO) systems employing three and four transmit antennas. We fill the empty slots of the existing STBCs from CIODs in their transmission matrices by additional symbols and use the conditional ML decoding technique which significantly reduces the ML decoding complexity of non-orthogonal STBCs while ensuring full-diversity and high coding gain. First, two new schemes with code rates of 2 and 1.5 are proposed for MIMO systems with four transmit antennas. We show that our low-complexity rate-2 STBC outperforms the corresponding best STBC recently proposed by Biglieri et al. for QPSK, due to its superior coding gain while our rate-1.5 STBC outperforms the full-diversity quasi-orthogonal STBC (QOSTBC). Then, two STBCs with code rates of 2 and 1.5 are proposed for three transmit antennas which are shown to outperform the corresponding full-diversity QOSTBC for three transmit antennas. We prove by an information-theoretic analysis that the capacities of new rate-2 STBCs for three and four transmit antennas are much closer to the actual MIMO channel capacity than the capacities of classical OSTBCs and CIODs.

# 1. INTRODUCTION

It has been shown that the capacity of wireless channels can be significantly increased by the use of multiple antennas [1]. Therefore, multiple-input multiple-output (MIMO) transmission techniques have attracted too much attention to realise the promising potential of multiple antennas. Space-time block codes (STBCs) offer an effective way to exploit this potential because of their simplicity and high performance. In 1998, Alamouti invented a remarkable scheme [2] for MIMO systems with two transmit antennas, which allows low-complexity maximum likelihood (ML) decoding due to its orthogonality. Orthogonal STBCs (OSTBCs), which allow symbol-wise decoding, are then generalised for three and four transmit antennas in [3]. For such codes, the total ML decoding complexity is linear and proportional to the size of the signal constellation since all symbols can be decoded independently from each other. Although OSTBCs can be decoded with minimum decoder complexity, the orthogonality constraint is too restrictive. Moreover, in [4], it has been proved that the code rate of an OSTBC is upper bounded by 3/4 transmitted symbols per channel use for more than two transmit antennas. Related by this bound in transmission code rate, from an information-theoretic point of view, OSTBCs can cause a significant loss in MIMO channel capacity [5]. Therefore, researchers have focused on increasing the code rates of STBCs by relaxing orthogonality constraint. Quasi-orthogonal STBCs (QOSTBCs), which exceed the upper bound mentioned above with a higher decoding complexity, have been proposed for three and four transmit antennas [6, 7]. These original schemes are then improved to obtain full-diversity by rotating some of the information symbols [8, 9]. STBCs using coordinate interleaved orthogonal designs (CIODs) proposed in [10] allow single-symbol decoding which enables easy ML decoders and offer higher data rates than OSTBCs for three and four transmit antennas. However, since years, the demand for STBCs with higher data rates has not ceased since symbol rate 1 may not be sufficient for next generation wireless communication systems [11]. One way to obtain full-diversity STBCs with higher data rates is to use algebraic number theory and cyclic division algebras; however these algebraic codes have very-high ML decoding complexities. A well-known example to such codes for two transmit antennas is the rate-2 Golden Code [12], which is reported to have a decoding complexity that grows with the fourth power of the constellation size. Two alternative STBCs are recently proposed by Parades et al. [13] and Sezginer and Sari [14] with lower decoding complexity and a slight degradation in error performance. These STBCs have a ML decoding complexity that is proportional with the third power of the constellation size. For four transmit antennas, the best known scheme was known as the DjABBA code [11, 15], however recently Biglieri, Hong and Viterbo (BHV) proposed the scheme in [16] which is reported to outperform all existing schemes for QPSK. However, when compared with OSTBCs and QOSTBCs, both of the STBCs in [15] and [16] have a very-high decoding complexity which is proportional to the seventh power of the constellation size. A rate-1.5 STBC has been proposed for four transmit antennas in [17] which has an identical error performance with the QOSTBC in [7, 9]. To the best of our knowledge, there is no rate-2 or rate-1.5 STBC for three transmit antennas given in literature.

This paper deals with the design of low ML decoding complexity rate-2 and rate-1.5 full-diversity STBCs for three and four transmit antennas. To reduce the decoding complexity of these codes, we use the conditional ML decoding technique, recently used for decoding of the non-orthogonal STBC in [14]. For four transmit antennas, we propose a new rate-2 STBC that achieves better error performance with a lower decoding complexity than the BHV code [16] for QPSK due to its higher coding gain. Moreover, a new rate-1.5 STBC is proposed for four transmit antennas which outperforms the STBCs in [9] and [17] for QPSK. Finally, two STBCs with rates of 2 and 1.5 are proposed for three transmit antennas which are shown to outperform the corresponding full-diversity QOSTBC. These better error performances of the proposed STBCs are the result of their optimised design parameters for QPSK constellation. An information-theoretic analysis is performed for the new rate-2 STBCs which shows that when compared with the OSTBCs, the new rate-2 STBCs maximise the potential of multiple antennas in terms of ergodic channel capacity.

The rest of the paper can be summarised as follows. We give our channel model and design criteria in Section 2. In Section 3, we review the conditional ML decoding technique and demonstrate the way we start our discussion. In Sections 4 and 5 we introduce the high-rate STBCs for four and three transmit antennas, respectively. Information-theoretic analysis for the proposed rate-2 STBCs is given in Section 6. We give performance comparisons in Section 7 and our conclusions in Section 8.

## 2. CHANNEL MODEL AND CODE DESIGN CRITERIA

Let us consider an $n_T \times n_R$ quasi-static Rayleigh flat fading MIMO channel, where $n_T$ and $n_R$ denote the number of transmit and receive antennas, respectively. The received $T \times n_R$ signal matrix $\mathbf{Y} \in \mathbb{C}^{T \times n_R}$ can be modeled as

$$\mathbf{Y} = \mathbf{XH} + \mathbf{N} \qquad (1)$$

where $\mathbf{X} \in \mathbb{C}^{T \times n_T}$ is the codeword (transmission) matrix, transmitted over $T$ channel uses. $\mathbf{H}$ and $\mathbf{N}$ are the $n_T \times n_R$ channel matrix and the $T \times n_R$ noise matrix, respectively. The entries of $\mathbf{H}$ and $\mathbf{N}$ are i.i.d. complex Gaussian random variables with the pdfs $N_{\mathbb{C}}(0,1)$ and $N_{\mathbb{C}}(0, N_0)$, respectively. We assume, $\mathbf{H}$ remains constant during the transmission of a codeword, and take independent values from one codeword to another. The realisation of $\mathbf{H}$ is assumed to be known at the receiver, but not at the transmitter. We give the following definitions:

*Definition 1*: (Code Rate) The code rate of a STBC with the codeword matrix **X** is defined as $R = k/T$ symbols per channel use where $k$ is the number of information symbols embedded in **X**. A STBC is said to be full-rate or high-rate if $R = 1$ or $R > 1$, respectively.

*Definition 2*: (Decoding Complexity) The ML decoding complexity is the number of metric computations performed to decode the codeword **X**.

By direct approach, ML decoding of **X** is performed by deciding in favour of the codeword which minimises the following metric

$$\hat{\mathbf{X}} = \arg\min_{\mathbf{X}} \|\mathbf{Y} - \mathbf{X}\mathbf{H}\|^2 \qquad (2)$$

where $\|.\|$ denotes the Frobenius norm. For a signal constellation of size $M$, the minimisation in (2) requires the computation of $M^k$ metrics which is the worst-case detection complexity since all the symbols in **X** are detected jointly. Note that OSTBCs [2, 3, 10] allow the decomposition of (2) to $k$ individual metrics each having a complexity of $M$, i.e., a total decoding complexity of $kM$ is obtained. A non-orthogonal STBC said to be reduced complexity if its ML detection is performed with less that $M^k$ total metric computations.

*Definition 3*: (Full-Diversity STBC) Let $r$ denote the rank of the codeword difference matrix $(\mathbf{X} - \hat{\mathbf{X}})$, with $\mathbf{X} \neq \hat{\mathbf{X}}$. A STBC is said to be full-diversity if $(\mathbf{X} - \hat{\mathbf{X}})$ is full-rank for all realisations of the possible codeword pairs. In this case, $r = n_T$, and the resulting diversity gain is $n_T n_R$ at high SNR.

For a full-diversity STBC, the worst-case pairwise error probability (PEP) also depends asymptotically to the minimum determinant $\delta_{\min}$, defined as

$$\delta_{\min} = \min_{\mathbf{X} \neq \hat{\mathbf{X}}} \det\left[(\mathbf{X} - \hat{\mathbf{X}})(\mathbf{X} - \hat{\mathbf{X}})^H\right] \qquad (3)$$

where $(.)^H$ denotes Hermitian transpose, the resulting coding gain being $(\delta_{\min})^{1/n_t}$. The rank and determinant criteria [18] provide the maximisation of diversity and coding gains. Note that for high signal-to-noise ratio (SNR), the dominant parameter is the diversity gain which determines the slope of the error curve. After the full-diversity is ensured, we have to maximise $\delta_{min}$ to obtain optimum performance.

### 3. DESIGN PROCEDURE AND CONDITIONAL ML DECODING

Let $\mathbf{Q}_{n,k}$ denotes an OSTBC for *n* transmit antennas such those given in [10], which transmits *k* information symbols $(x_1, x_2, ..., x_k)$ with having empty slots left in its codeword matrix for orthogonality, we obtain *k+λ* information symbols transmitting high-rate, full-diversity STBC $\mathbf{X}_{n,k+\lambda}$ from $\mathbf{Q}_{n,k}$ as

$$\mathbf{X}_{n,k+\lambda} = \mathbf{Q}_{n,k} + \mathbf{PG}_\lambda \qquad (4)$$

where $\mathbf{G}_\lambda$ is the codeword matrix with $\lambda$ additional information symbols to be transmitted from empty slots of $\mathbf{Q}_{n,k}$. $\mathbf{P}$ is the optimisation matrix whose entries are complex design parameters to be determined by the rank and determinant criteria. $\mathbf{Q}_{n,k}$ and $\mathbf{PG}_\lambda$ contain non-overlapping entries. Due to non-orthogonal structure of $\mathbf{X}_{n,k+\lambda}$, by direct computation, $M^{(k+\lambda)}$ metric computations are required for the ML decoding, i.e.,

$$\hat{\mathbf{X}}_{n,k+\lambda} = \arg \min_{x_1, x_2, ..., x_{k+\lambda}} \left\| \mathbf{Y} - \mathbf{X}_{n,k+\lambda} \mathbf{H} \right\|^2. \qquad (5)$$

When compared with the decoding complexity of $\mathbf{Q}_{n,k}$, which is *kM*, this increase in complexity is unacceptable. However, we try to eliminate in (5) the terms coming from additional transmitted symbols from empty slots of $\mathbf{Q}_{n,k}$, by computing intermediate signals from the received signals for all possible values of the additional symbols $x_{k+1}, x_{k+2}, ..., x_{k+\lambda}$ in $\mathbf{G}_\lambda$, as

$$\mathbf{Z} = \mathbf{Y} - \mathbf{PG}_\lambda \mathbf{H}. \qquad (6)$$

Going over this search for all combinations of $x_{k+1}, x_{k+2},...,x_{k+\lambda}$, we use the decoding procedure of $\mathbf{Q}_{n,k}$ to obtain conditional ML estimates of $x_1, x_2,...,x_k$ given $x_{k+1}, x_{k+2},...,x_{k+\lambda}$, although only for the correct combination of $x_{k+1}, x_{k+2},...,x_{k+\lambda}$ (6) reduces to

$$\mathbf{Z} = \mathbf{Q}_{n,k}\mathbf{H} + \mathbf{N}. \tag{7}$$

Finally, we minimise the decision metric given in (5) for $x_1^{ML}, x_2^{ML},...,x_k^{ML}, x_{k+1}, x_{k+2},...,x_{k+\lambda}$ over all possible values of $x_{k+1}, x_{k+2},...,x_{k+\lambda}$. In other words, instead of searching over all possible values of $x_1, x_2,...,x_{k+\lambda}$ and suffering from $M^{k+\lambda}$ metric computations, we only search with a decoding complexity of $M^\lambda$, and obtain conditional ML estimates of $x_1, x_2,...,x_k$, which needs an additional decoding complexity of $kM$ per each step of $M^\lambda$ calculations. Therefore, we obtain a total decoding complexity of $kM \times M^\lambda = kM^{\lambda+1}$. Empirical tests show that the use of conditional ML technique, gives the same result with the direct approach given in (5).

## 4. NEW RATE-2 AND RATE-1.5 STBCS FOR FOUR TRANSMIT ANTENNAS

In this section we propose rate-2 and rate-1.5 STBCs for four transmit antennas by using the high-rate STBC design procedure given in Section 3. Let us consider the rate-1 CIOD for four transmit antennas [10], which takes a block of four modulated symbols and transmits them from four antennas in four time intervals according to the code matrix given by

$$\mathbf{Q}_{4,4} = \begin{bmatrix} x_{0R} + jx_{2I} & x_{1R} + jx_{3I} & 0 & 0 \\ -(x_{1R} + jx_{3I})^* & (x_{0R} + jx_{2I})^* & 0 & 0 \\ 0 & 0 & x_{2R} + jx_{0I} & x_{3R} + jx_{1I} \\ 0 & 0 & -(x_{3R} + jx_{1I})^* & (x_{2R} + jx_{0I})^* \end{bmatrix} \tag{8}$$

where $x_{iR}$ and $x_{iI}$ for $i = 0,...,3$ denote real and imaginary parts of $x_i$, respectively. According to (4), we propose the following rate-2 STBC which transmits eight information symbols in four time intervals,

$$\mathbf{X}_{4,8} = \begin{bmatrix} x_{0R} + jx_{2I} & x_{1R} + jx_{3I} & e^{j\theta}(x_{4R} + jx_{6I}) & e^{j\theta}(x_{5R} + jx_{7I}) \\ -(x_{1R} + jx_{3I})^* & (x_{0R} + jx_{2I})^* & -e^{j\theta}(x_{5R} + jx_{7I})^* & e^{j\theta}(x_{4R} + jx_{6I})^* \\ x_{6R} + jx_{4I} & x_{7R} + jx_{5I} & x_{2R} + jx_{0I} & x_{3R} + jx_{1I} \\ -(x_{7R} + jx_{5I})^* & (x_{6R} + jx_{4I})^* & -(x_{3R} + jx_{1I})^* & (x_{2R} + jx_{0I})^* \end{bmatrix} \quad (9)$$

for the optimisation matrix

$$\mathbf{P} = \begin{bmatrix} e^{j\theta} & 0 & 0 & 0 \\ 0 & e^{j\theta} & 0 & 0 \\ 0 & 0 & 1 & 0 \\ 0 & 0 & 0 & 1 \end{bmatrix}. \quad (10)$$

An exhaustive computer search was performed for unit energy QPSK signal constellation to check the non-vanishing determinant property for $\mathbf{X}_{4,8}$. For $\theta = 90°$, we obtained the maximum $\delta_{min}$ value of 0.64 which corresponds to that for $\mathbf{Q}_{4,4}$ for the same average transmitted signal energy per symbol. Therefore, we conclude that the matrix in (10) is optimum for $\theta = 90°$ in terms of coding gain. Note that, for the coordinate interleaved STBCs in (8) and (9), the QPSK signal constellation with symbols on the two axes must be rotated by an angle of 13.29° to ensure full-diversity and maximum coding gain [10].

The decoding procedure for $\mathbf{X}_{4,8}$ is given as follows. The receiver calculates intermediate signals from the received signals for all possible values of $x_4, x_5, x_6$ and $x_7$, and since for the correct combination, intermediate signals are found as

$$\mathbf{Z} = \mathbf{Q}_{4,4}\mathbf{H} + \mathbf{N} \quad (11)$$

the receiver follows the decoding procedure of $\mathbf{Q}_{4,4}$. Let $z_{ij} \in \mathbf{Z}$ be the intermediate signal calculated from $r_{ij} \in \mathbf{Y}$, $i$ and $j$ denoting the $i$th column and $j$th row of the corresponding matrix. The receiver combines the intermediate signals to obtain $\tilde{y}_0 = \sum_{i=1}^{n_R}\left(h_{i,1}^* z_{i,1} + h_{i,2} z_{i,2}^*\right)$, $\tilde{y}_1 = \sum_{i=1}^{n_R}\left(h_{i,2}^* z_{i,1} - h_{i,1} z_{i,2}^*\right)$, $\tilde{y}_2 = \sum_{i=1}^{n_R}\left(h_{i,3}^* z_{i,3} + h_{i,4} z_{i,4}^*\right)$, and $\tilde{y}_3 = \sum_{i=1}^{n_R}\left(h_{i,4}^* z_{i,3} - h_{i,3} z_{i,4}^*\right)$, then uses the following rules to obtain ML estimates for $x_i$, $i=0,\ldots,3$ conditioned on the quadruple $x_4, x_5, x_6$ and $x_7$,

$$\begin{aligned} x_0^{ML} &= \arg\min_{x_0} \left\{ \left(\beta|\hat{x}_{0R} - \alpha x_{0R}|^2\right) + \left(\alpha|\hat{x}_{0I} - \beta x_{0I}|^2\right) \right\} \\ x_1^{ML} &= \arg\min_{x_1} \left\{ \left(\beta|\hat{x}_{1R} - \alpha x_{1R}|^2\right) + \left(\alpha|\hat{x}_{1I} - \beta x_{1I}|^2\right) \right\} \\ x_2^{ML} &= \arg\min_{x_2} \left\{ \left(\alpha|\hat{x}_{2R} - \beta x_{2R}|^2\right) + \left(\beta|\hat{x}_{2I} - \alpha x_{2I}|^2\right) \right\} \\ x_3^{ML} &= \arg\min_{x_3} \left\{ \left(\alpha|\hat{x}_{3R} - \beta x_{3R}|^2\right) + \left(\beta|\hat{x}_{3I} - \alpha x_{3I}|^2\right) \right\} \end{aligned} \quad (12)$$

where $\alpha = \sum_{i=1}^{n_R}\left(|h_{i,1}|^2 + |h_{i,2}|^2\right)$, $\beta = \sum_{i=1}^{n_R}\left(|h_{i,3}|^2 + |h_{i,4}|^2\right)$ and $\hat{x}_0 = \operatorname{Re}\{\tilde{y}_0\} + j\operatorname{Im}\{\tilde{y}_2\}$, $\hat{x}_1 = \operatorname{Re}\{\tilde{y}_1\} + j\operatorname{Im}\{\tilde{y}_3\}$, $\hat{x}_2 = \operatorname{Re}\{\tilde{y}_2\} + j\operatorname{Im}\{\tilde{y}_0\}$, $\hat{x}_3 = \operatorname{Re}\{\tilde{y}_3\} + j\operatorname{Im}\{\tilde{y}_1\}$. According to the conditional ML decoding procedure given in Section 3, for $k = \lambda = 4$ we obtain a total decoding complexity of $4M^5$ instead of $M^8$ by minimizing (5) for $x_0^{ML}, x_1^{ML}, x_2^{ML}, x_3^{ML}, x_4, x_5, x_6, x_7$ over all possible values of $x_4, x_5, x_6$ and $x_7$.

For a further reduction in ML decoding complexity, (9) can be modified by setting $x_6 = x_7 = 0$ [†] to obtain a new rate-1.5 STBC, given as

$$\mathbf{X}_{4,6} = \begin{bmatrix} x_{0R} + jx_{2I} & x_{1R} + jx_{3I} & e^{j\theta}(x_{4R}) & e^{j\theta}(x_{5R}) \\ -(x_{1R} + jx_{3I})^* & (x_{0R} + jx_{2I})^* & -e^{j\theta}(x_{5R}) & e^{j\theta}(x_{4R}) \\ jx_{4I} & jx_{5I} & x_{2R} + jx_{0I} & x_{3R} + jx_{1I} \\ -(jx_{5I})^* & (jx_{4I})^* & -(x_{3R} + jx_{1I})^* & (x_{2R} + jx_{0I})^* \end{bmatrix}. \quad (13)$$

Since the optimisation matrix of $\mathbf{X}_{4,8}$ is optimum for $\theta = 90°$ in terms of $\delta_{\min}$, by using the same optimisation matrix for $\mathbf{X}_{4,6}$, we obtain full-diversity with the maximum possible $\delta_{\min}$ value of 0.64. The decoding of $\mathbf{X}_{4,6}$ is similar to that of $\mathbf{X}_{4,8}$. By calculating intermediate signals for all possible values of $x_4$ and $x_5$, the receiver obtains $\mathbf{Z}$ from (11), and following the same decoding procedures in (12), it obtains ML estimates for $x_i$, $i = 0,...,3$ conditioned on the pair $(x_4, x_5)$. Instead of $M^6$, a total decoding complexity of $4M^3$ is obtained since $k = 4$ and $\lambda = 2$.

---

[†] A rate-7/4 STBC $\mathbf{X}_{4,7}$ is also possible by setting only $x_7 = 0$ in (9). Since the optimisation matrix in (10) is optimum for $\mathbf{X}_{4,8}$ when $\theta = 90°$, the same $\delta_{\min}$ value of 0.64 is obtained for $\mathbf{X}_{4,7}$. A total decoding complexity of $4M^4$ is required to decode $\mathbf{X}_{4,7}$ since $k = 4$ and $\lambda = 3$.

## 5. NEW RATE-2 AND RATE-1.5 STBCS FOR THREE TRANSMIT ANTENNAS

In this section we propose two novel STBCs with rates of 2 and 1.5 for three transmit antennas. Let us consider the generalised CIOD for three transmit antennas from [10],

$$\mathbf{Q}_{3,4} = \begin{bmatrix} x_{0R} + jx_{2I} & x_{1R} + jx_{3I} & 0 \\ -(x_{1R} + jx_{3I})^* & (x_{0R} + jx_{2I})^* & 0 \\ 0 & 0 & x_{2R} + jx_{0I} \\ 0 & 0 & -(x_{3R} + jx_{1I})^* \end{bmatrix}. \tag{14}$$

According to (4), we obtain a rate-2, full-diversity STBC as

$$\mathbf{X}_{3,8} = \begin{bmatrix} x_{0R} + jx_{2I} & x_{1R} + jx_{3I} & e^{j\theta}(x_{4R} + jx_{6I}) \\ -(x_{1R} + jx_{3I})^* & (x_{0R} + jx_{2I})^* & -e^{j\theta}(x_{5R} + jx_{7I})^* \\ e^{j\theta}(x_{6R} + jx_{4I}) & e^{j\theta}(x_{7R} + jx_{5I}) & x_{2R} + jx_{0I} \\ -e^{j\theta}(x_{7R} + jx_{5I})^* & e^{j\theta}(x_{6R} + jx_{4I})^* & -(x_{3R} + jx_{1I})^* \end{bmatrix} \tag{15}$$

for the optimisation matrix $\mathbf{P} = e^{j\theta}\mathbf{I}_4$ where $\mathbf{I}_4$ is the $4 \times 4$ identity matrix. An exhaustive computer search was performed to obtain maximum coding gain for $\mathbf{X}_{3,8}$ by optimising $\theta$. The optimum value for $\theta$ was found as $13.91°$ which gives a $\delta_{min}$ value of $0.1564$[‡] for QPSK while the $\delta_{min}$ of $\mathbf{Q}_{3,4}$ is equal to $0.3381$. Note that, the optimum constellation rotation angle for the STBCs in (14) and (15) is equal to $16°$ for QPSK. $\mathbf{X}_{3,8}$ is decoded with the same manner as $\mathbf{X}_{4,8}$ by taking $h_{i,4} = 0, i = 1,...,n_R$ for combining and ML decision rules. Similar to four transmit antennas case, we obtain a total decoding complexity of $4M^5$ instead of $M^8$ since $k = \lambda = 4$.

---

[‡]Since the codeword matrices for $\mathbf{Q}_{3,4}$ and $\mathbf{X}_{3,8}$ are non-square, the determinant of the codeword difference matrix of $\mathbf{Q}_{3,4}$ and $\mathbf{X}_{3,8}$ is zero while its rank is 3, i.e. full since $r = n_T$. In this case we calculate the minimum determinant as $\delta_{min} = \min_{\mathbf{X} \neq \hat{\mathbf{X}}} \prod_{i=1}^{3} \lambda_i$ where, $\lambda_i$ is the non-zero eigenvalues of the distance matrix $(\mathbf{X} - \hat{\mathbf{X}})(\mathbf{X} - \hat{\mathbf{X}})^H$.

By modifying (15), a rate-1.5 STBC, which transmits six information symbols at four time intervals, is obtained as follows

$$\mathbf{X}_{3,6} = \begin{bmatrix} x_{0R} + jx_{2I} & x_{1R} + jx_{3I} & e^{j\theta}(x_{4R}) \\ -(x_{1R} + jx_{3I})^* & (x_{0R} + jx_{2I})^* & -e^{j\theta}(x_{5R}) \\ e^{j\theta}(jx_{4I}) & e^{j\theta}(jx_{5I}) & x_{2R} + jx_{0I} \\ -e^{j\theta}(jx_{5I})^* & e^{j\theta}(jx_{4I})^* & -(x_{3R} + jx_{1I})^* \end{bmatrix} \quad (16)$$

for the optimisation matrix $\mathbf{P} = e^{j\theta}\mathbf{I}_4$. Unlike $\mathbf{X}_{3,8}$, for $\theta = 45°$, we obtained the same $\delta_{\min}$ value as for $\mathbf{Q}_{3,4}$, which is equal to 0.3381. Therefore the maximum possible $\delta_{\min}$ value is achieved for $\mathbf{X}_{3,6}$ while the total decoding complexity is reduced from $M^6$ to $4M^3$ for $k = 4$ and $\lambda = 2$.

## 6. INFORMATION-THEORETIC ANALYSIS OF NEW RATE-2 STBCS

In this section, we analyze the maximum mutual information (MMI) achieved by our rate-2 STBC designs given in previous sections and compare them with MMI achieved by classical CIODs and the actual MIMO channel capacity. We start by the ergodic capacity of a $n_T \times n_R$ MIMO channel [1], which is characterised by a $n_R \times n_T$ channel matrix $\mathbf{H}$ that is known at the receiver but not at the transmitter. At an SNR value $\rho$, the ergodic MIMO capacity is given as

$$C(\rho, n_T, n_R) = E\left\{\log \det\left(\mathbf{I}_{n_T} + \frac{\rho}{n_T}\mathbf{H}^H\mathbf{H}\right)\right\} \quad (17)$$

where the expectation is taken over the distribution of the random channel matrix $\mathbf{H}$. To perform an information-theoretic analysis, the channel model in (1) must be modifed as

$$\mathbf{y} = \sqrt{\frac{\rho}{n_T}}H\mathbf{x} + \mathbf{n} \quad (18)$$

where $H$ is the *equivalent channel matrix* [5] of the STBC **X**, **y**, **x** and **n** are the received signal, unit-variance transmitted signal and noise vectors, respectively. The normalisation factor in (18) ensures that $\rho$ is the SNR at each receive antenna. For the CIOD $\mathbf{Q}_{4,4}$ given in (8), the equivalent channel model with $n_R$ receive antennas can be expressed from (18) as

$$\mathbf{y} = \sqrt{\frac{\rho}{4}}\sqrt{2}\underbrace{\begin{bmatrix} H_1 \\ H_2 \\ \vdots \\ H_{n_R} \end{bmatrix}}_{H_{4n_R \times 4}} \underbrace{\begin{bmatrix} x_{0R} + jx_{2I} \\ x_{1R} + jx_{3I} \\ x_{2R} + jx_{0I} \\ x_{3R} + jx_{1I} \end{bmatrix}}_{\mathbf{x}} + \mathbf{n}. \qquad (19)$$

where

$$H_l = \begin{bmatrix} h_{l,1} & h_{l,2} & 0 & 0 \\ h_{l,2}^* & -h_{l,1}^* & 0 & 0 \\ 0 & 0 & h_{l,3} & h_{l,4} \\ 0 & 0 & h_{l,4}^* & -h_{l,3}^* \end{bmatrix} \quad \text{with } l = 1,\ldots,n_R$$

and $H_{4n_R \times 4}$ is the $4n_R \times 4$ equivalent channel matrix for $\mathbf{Q}_{4,4}$. The MMI attained by $\mathbf{Q}_{4,4}$ is given as [10]

$$\begin{aligned}
C_{\mathbf{Q}_{4,4}}(\rho, 4, n_R) &= \frac{1}{4} E\left\{\log\det\left(\mathbf{I}_4 + \frac{\rho}{4} H_{4n_R \times 4}^H H_{4n_R \times 4}\right)\right\} \\
&= \frac{1}{2} E\left\{\log\left(1 + \frac{\rho}{2}\sum_{i=1}^{n_R}\left[|h_{i,1}|^2 + |h_{i,2}|^2\right]\right)\right\} + \frac{1}{2} E\left\{\log\left(1 + \frac{\rho}{2}\sum_{i=1}^{n_R}\left[|h_{i,3}|^2 + |h_{i,4}|^2\right]\right)\right\} \quad (20) \\
&= \frac{1}{2} C(n_R\rho, 2n_R, 1) + \frac{1}{2} C(n_R\rho, 2n_R, 1) = C(n_R\rho, 2n_R, 1) < C(\rho, 4, n_R)
\end{aligned}$$

where the factor 1/4 normalises for the four channel uses spanned by $\mathbf{Q}_{4,4}$. We conclude that $\mathbf{Q}_{4,4}$ can not achieve full channel capacity even for $n_R = 1$ which can be explained by the zeros in (8). On the other hand, for $\mathbf{X}_{4,8}$ the equivalent channel model with $n_R$ receive antennas is given from (18) as

$$\mathbf{y} = \sqrt{\frac{\rho}{4}} \underbrace{\begin{bmatrix} H_1 \\ H_2 \\ \vdots \\ H_{n_R} \end{bmatrix}}_{H_{4n_R \times 8}} \underbrace{\begin{bmatrix} x_{0R} + jx_{2I} \\ x_{1R} + jx_{3I} \\ x_{2R} + jx_{0I} \\ x_{3R} + jx_{1I} \\ x_{4R} + jx_{6I} \\ x_{5R} + jx_{7I} \\ x_{6R} + jx_{4I} \\ x_{7R} + jx_{5I} \end{bmatrix}}_{\mathbf{x}} + \mathbf{n} \tag{21}$$

where

$$H_l = \begin{bmatrix} h_{l,1} & h_{l,2} & 0 & 0 & jh_{l,3} & jh_{l,4} & 0 & 0 \\ h_{l,2}^* & -h_{l,1}^* & 0 & 0 & -jh_{l,4}^* & jh_{l,3}^* & 0 & 0 \\ 0 & 0 & h_{l,3} & h_{l,4} & 0 & 0 & h_{l,1} & h_{l,2} \\ 0 & 0 & h_{l,4}^* & -h_{l,3}^* & 0 & 0 & h_{l,2}^* & -h_{l,1}^* \end{bmatrix} \quad \text{with } l = 1,\ldots,n_R$$

and $H_{4n_R \times 8}$ is the $4n_R \times 8$ equivalent channel matrix for $\mathbf{X}_{4,8}$. The MMI of the new rate-2 STBC $\mathbf{X}_{4,8}$ is obtained as

$$C_{\mathbf{X}_{4,8}}(\rho, 4, n_R) = \frac{1}{4} E\left\{ \log \det\left( \mathbf{I}_8 + \frac{\rho}{4} H_{4n_R \times 8}^H H_{4n_R \times 8} \right) \right\}. \tag{22}$$

Due to the complexity of determinant calculations for (22), $C_{\mathbf{X}_{4,8}}(\rho, 4, n_R)$ is directly evaluated by Monte-Carlo simulations. For three transmit antennas, after appropriate normalisations in (18), the MMI attained by $\mathbf{Q}_{3,4}$ is calculated as

$$C_{\mathbf{Q}_{3,4}}(\rho, 3, n_R) = \frac{1}{2}\left[ C\left( \frac{4\rho n_R}{3}, 2n_R, 1 \right) + C\left( \frac{2\rho n_R}{3}, n_R, 1 \right) \right] < C(\rho, 3, n_R). \tag{23}$$

Finally, we obtain the MMI attained by the new rate-2 STBC $\mathbf{X}_{3,8}$ as

$$C_{\mathbf{X}_{3,8}}(\rho, 3, n_R) = \frac{1}{4} E\left\{ \log \det\left( \mathbf{I}_8 + \frac{\rho}{3} H_{4n_R \times 8}^H H_{4n_R \times 8} \right) \right\} \tag{24}$$

where $H_{4n_R \times 8}$ is the $4n_R \times 8$ equivalent channel matrix for $\mathbf{X}_{3,8}$. It is shown in [10] that the capacities of $\mathbf{Q}_{4,4}$ and $\mathbf{Q}_{3,4}$ are greater than those of rate-3/4 OSTBCs[§] for three and four transmit antennas, respectively. However, the zeros in $\mathbf{Q}_{4,4}$ and $\mathbf{Q}_{3,4}$ prevent them achieving the actual MIMO channel capacity even for one receive antenna. In Figs. 1-2, the MMI of rate-1 CIODs and rate-2 STBCs are depicted for four and three transmit, one and two receive antenna cases. As seen from Figs. 1-2, for one receive antenna case, both $\mathbf{X}_{4,8}$ and $\mathbf{X}_{3,8}$ achieve the actual channel capacity, however, when the number of receive antennas are increased to two, they suffer a slight loss. On the other hand, while the capacity loss of the orthogonal designs is negligible for one receive antenna, this loss becomes substantial for more than one receive antenna since these schemes have lower transmission rates compared to the proposed STBCs.

## 7. SIMULATION RESULTS AND COMPARISONS

In this section, we evaluate the bit error rate (BER) performance of the proposed STBCs by computer simulations and compare the results with the existing comparable schemes given in the literature. Bit error rate (BER) curves of the proposed STBC $\mathbf{X}_{4,8}$ and the BHV code [16] for a 4×2 MIMO system operating on a quasi-static Rayleigh fading channel are depicted in Fig. 3 as a function of received SNR for QPSK constellation corresponding to a transmission data rate of 4 bits/s/Hz for both schemes. From these curves, we conclude that the new STBC achieves better error performance than the BHV code and the performance gap between the BHV code and the new STBC increases with increasing SNR values due to the diversity loss of the BHV code since its $\delta_{\min}$ value is zero. In [16], the BHV code is reported to outperform all existing rate-2 schemes for four transmit antennas. Until [16], the best known rate-2 STBC for four transmit antennas was known as the DjABBA code [11, 15], whose $\delta_{min}$ value is equal to 0.04 for the same average total transmitted power with $\mathbf{X}_{4,8}$. However, the better performance of the BHV code is explained in [16] by the optimisation of its multiplicity defined as the total number of different codeword pairs giving $\delta_{min}$. BER performance of the QOSTBC [7, 9] for 16-QAM is omitted, since it performs significantly worse (approximately 2.5 dB) than rate-2 STBCs. BER performance of the proposed STBC $\mathbf{X}_{3,8}$ is also depicted in Fig. 3 and compared with the best known STBC for three transmit antennas which is the full-diversity QOSTBC with constellation rotation [9], obtained by removing the last column of the QOSTBC in [6]. Our STBC uses QPSK while QOSTBC uses 16-QAM, i.e., both schemes have a transmission rate of 4 bit/s/Hz. Approximately, 1.4dB SNR advantage is obtained by the new scheme which provides an increase in code rate by a factor of 2 while ensuring full-diversity and high coding gain.

---

[§] It is shown in [19] that the capacity of a rate-$R$ OSTBC for $n_T$ transmit antennas is given as $C_{OSTBC}(\rho, n_T, n_R) = RC(\rho n_R / R, n_T n_R, 1)$ which is smaller than the capacity of CIODs for $n_T > 2$ since when $n_T > 2$, $R \leq 3/4$ for OSTBCs [4].

BER curves of the proposed rate-1.5 STBCs are given in Fig. 4. To obtain a transmission rate of 3 bits/s/Hz, rate-1.5 schemes use QPSK while reference rate-1 full-diversity QOSTBCs use non-rectangular 8-QAM for 4 and 3 transmit antennas. From these curves we conclude that our STBC $\mathbf{X}_{4,6}$ has an approximately 1.5 dB SNR advantage over QOSTBC of [7, 9]. For three transmit antennas, $\mathbf{X}_{3,6}$ provides approximately 0.6 dB SNR advantage over the QOSTBC of [6, 9]. These better performances are the result of an increase in code rate, since rate-1 QOSTBCs use larger and less efficient constellations with smaller normalised minimum Euclidean distance between symbols, than our rate-1.5 schemes, to achieve the same spectral efficiency. It should be noted that the performance gaps between the new rate-1.5 STBCs and QOSTBCs are lower than those between rate-2 STBCs and QOSTBCs. However, decoding complexity of rate-1.5 STBCs is $4M^3$ while decoding complexity of rate-2 STBCs is $4M^5$. Therefore, the proposed STBCs offer a trade-off between complexity and transmission rate.

## 8. CONCLUSIONS

We have derived an efficient method to obtain high-rate, full-diversity STBCs with simplified ML decoding, and applied to STBC designs from CIODs. We have shown that it is possible to obtain high-rate STBCs with significantly lower decoding complexities without degradation in error performance. A total of four schemes are proposed which offer a trade-off between code rate and complexity, and outperform their counterparts given in the literature in accordance with their optimised minimum determinants for QPSK constellation. Moreover, we have shown that the new rate-2 STBCs can more effectively exploit the potential of multiple antennas in terms of attainable capacity compared to the classical OSTBCs and CIODs. However, the optimisation of the proposed STBCs is left as a future work for higher constellations such as 16/64-QAM since minimum determinant searches take extremely long computation time.

**Figure Captions:**

Fig. 1: Maximum mutual information (ergodic) of new STBC ($\mathbf{X}_{4,8}$) and CIOD ($\mathbf{Q}_{4,4}$) for one and two receive antennas

Fig. 2: Maximum mutual information (ergodic) of new STBC ($\mathbf{X}_{3,8}$) and GCIOD ($\mathbf{Q}_{3,4}$) for one and two receive antennas

Fig. 3: BER performance of the proposed rate-2 STBCs for 4 bits/s/Hz

Fig. 4: BER performance of the proposed rate-1.5 STBCs for 3 bits/s/Hz

Fig.1

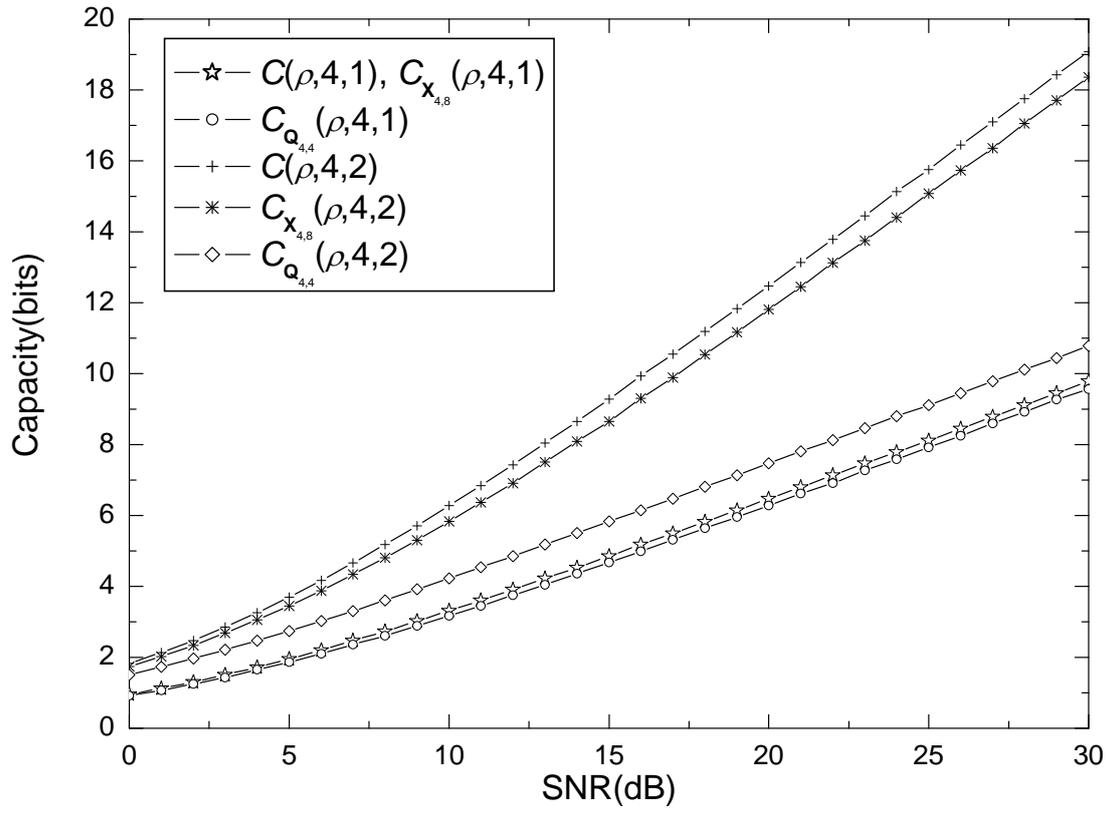

Fig.2

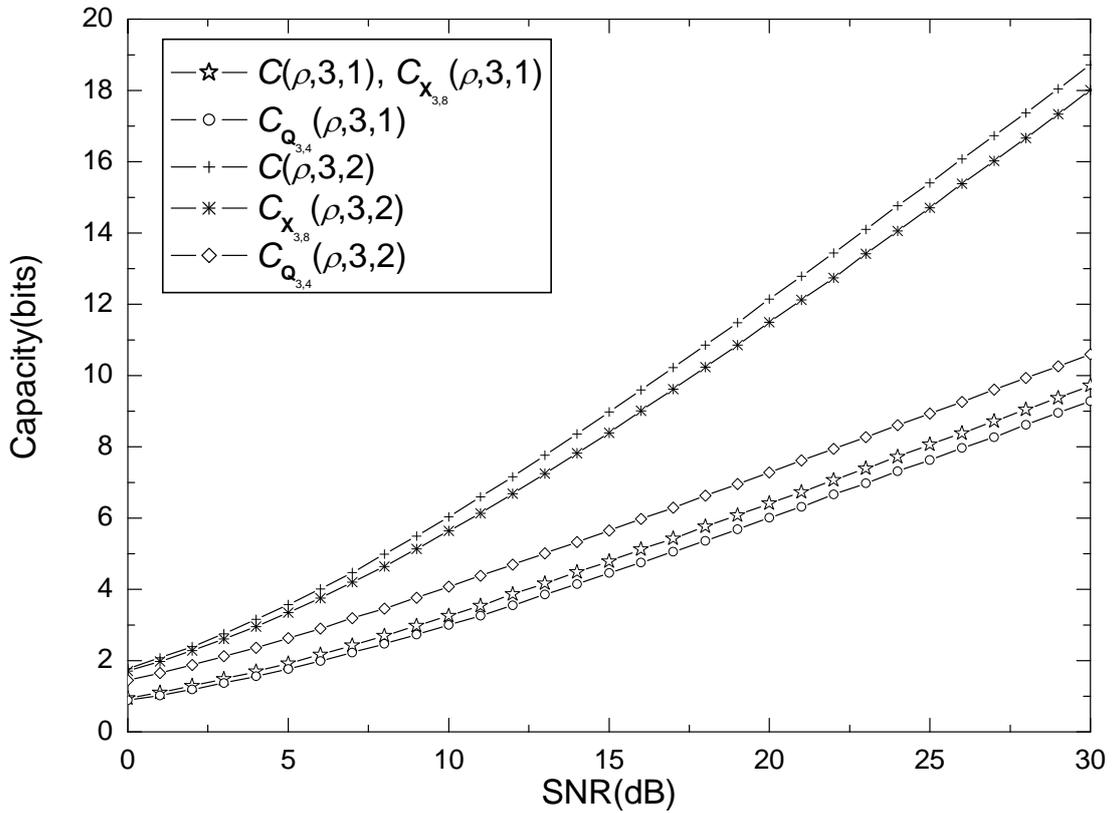

Fig. 3

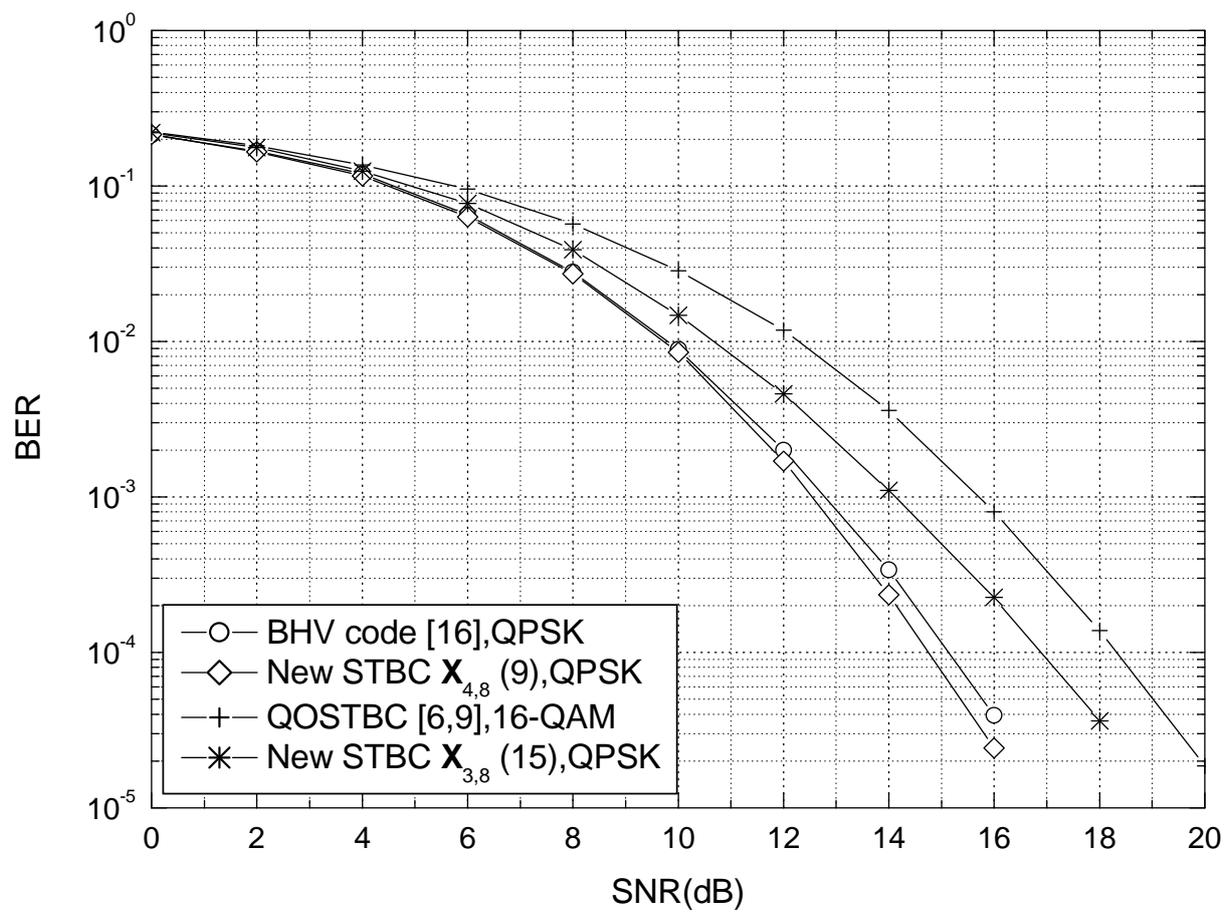

Fig. 4

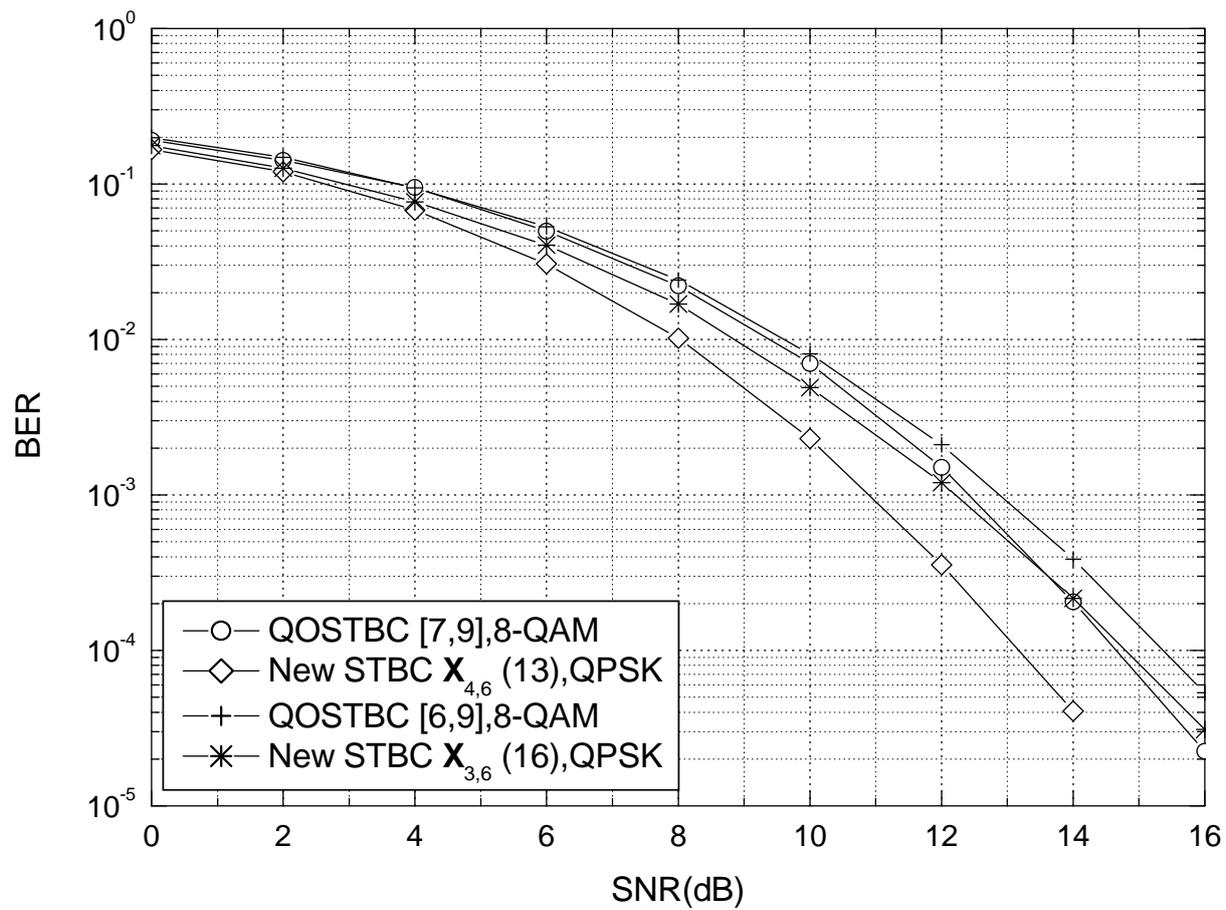